\documentclass[a4,12pt]{article}

\usepackage{graphicx}

\begin{document}

\begin{center}
{\Large\bf Direct Detection of the Cosmic Neutrino Background
Including Light Sterile Neutrinos}
\end{center}

\vspace{0.3cm}
\begin{center}
{\bf Y.F. Li} \footnote{E-mail: liyufeng@ihep.ac.cn}, 
~ {\bf Zhi-zhong Xing}
\footnote{E-mail: xingzz@ihep.ac.cn} \\
{\sl Institute of High Energy Physics and Theoretical Physics 
Center for Science Facilities, \\
Chinese Academy of Sciences, Beijing 100049, China}
\end{center}
\begin{center}
{\bf Shu Luo} \\
{\sl Department of Physics and Institute of Theoretical Physics and Astrophysics, Xiamen University, Xiamen, Fujian 361005, China}
\end{center}

\setcounter{footnote}{0}

\vspace{2.5cm}

\begin{abstract}
Current cosmological data drop an interesting hint about the
existence of sub-eV sterile neutrinos, which should be a part of the
cosmic neutrino background (C$\nu$B). We point out that such light
sterile neutrinos may leave a distinct imprint on the electron
energy spectrum in the capture of relic electron neutrinos by means
of radioactive beta-decaying nuclei. We examine possible signals of
sterile neutrinos relative to active neutrinos, characterized by
their masses and sensitive to their number densities, in the
reaction $\nu^{}_e + ~^3{\rm H} \to ~^3{\rm He} + e^-$ against the
corresponding tritium beta decay. We stress that this kind of direct
laboratory detection of the C$\nu$B and its sterile component might
not be hopeless in the long term.
\end{abstract}

\newpage

\framebox{\large\bf 1} ~ As fairly stable and weakly interacting
particles, relic neutrinos of the Big Bang must survive today and
form a cosmic background similar to the cosmic microwave background
(CMB) radiation. This cosmic neutrino background (C$\nu$B) played an
important role in the evolution of the Universe, and its existence
has been indirectly ``seen" from current cosmological data on the
Big Bang nucleosynthesis (BBN), CMB anisotropies and large-scale
structures \cite{PDG}. How to directly detect the C$\nu$B in a
laboratory experiment is a great challenge to the present
experimental techniques, simply because today's temperature of the
C$\nu$B is extremely low ($T^{}_\nu \approx 1.945$ K) and thus the
average three-momentum of each relic neutrino is very small
($\langle p^{}_\nu \rangle = 3T^{}_\nu \approx 5 \times 10^{-4}$
eV). Among several possibilities for the direct C$\nu$B detection
\cite{Ringwald}, the most promising one seems to be the relic
neutrino capture experiment by means of radioactive beta-decaying
nuclei \cite{Weinberg}---\cite{Kaboth}. The point is that a generic
neutrino capture reaction $\nu^{}_e + N \to N^\prime + e^-$ will
take place with no threshold on the incident neutrino energy,
provided $N$ can naturally undergo the beta decay $N \to N^\prime +
e^- + \overline{\nu}^{}_e$ with an energy release $Q^{}_\beta =
m^{}_N - m^{}_{N^\prime} - m^{}_e$ in the limit of vanishing
neutrino masses (i.e., $m^{}_i \to 0$ for $i=1,2,\cdots$). The
signal of this neutrino capture process is measured by the
monoenergetic electron's kinetic energy $Q^{}_\beta +
E^{}_{\nu^{}_i} \geq Q^{}_\beta + m^{}_i$ for each neutrino mass
eigenstate $\nu^{}_i$, as compared with the non-monoenergetic
electron's endpoint energy $Q^{}_\beta - m^{}_i$ for each $\nu^{}_i$
in the corresponding beta decay. So there is a gap equal to or
larger than $2m^{}_i$ between the kinetic energies of the detected
electrons in $\nu^{}_e + N \to N^\prime + e^-$ (signal) and $N \to
N^\prime + e^- + \overline{\nu}^{}_e$ (background). A measurement of
this gap will directly probe relic neutrinos and determine or
constrain their masses.

An immediate question is whether the C$\nu$B consists of only three
active neutrinos ($\nu^{}_e$, $\nu^{}_\mu$ and $\nu^{}_\tau$) or
more than three light neutrinos. Using $N^{}_{\rm eff}$ to denote
the effective number of thermally excited neutrino species in the
early Universe, some authors have recently obtained $N^{}_{\rm eff}
= 4.34^{+0.86}_{-0.88}$ at the $68\%$ confidence level from an
analysis of the 7-year WMAP data on CMB anisotropies and large-scale
structures \cite{WMAP10} or $N^{}_{\rm eff} = 4.78^{+1.86}_{-1.79}$
at the $95\%$ confidence level from a similar analysis including the
SDSS data on the (DR7) halo power spectrum \cite{Hamann1}. Moreover,
two independent groups have recently found slightly higher values of
the primordial $^4{\rm He}$ abundance \cite{BBN}, implying the
presence of additional radiation or relativistic particles during
the BBN epoch. As argued by Hamann {\it et al.} in Ref.
\cite{Hamann2}, these results drop an interesting hint that there
might exist one or more light sterile neutrinos in the C$\nu$B.
Their detailed analysis supports this conjecture and is compatible
with an interpretation of the LSND \cite {LSND} and MiniBOONE
\cite{MB} anomalies in terms of three active neutrinos and two
sterile neutrinos \cite{Karagiorgi} if the mass scale of sterile
neutrinos lies in the sub-eV range. Although such an interpretation
may have severe tension with current disappearance experiments of
neutrino oscillations \cite{Maltoni}, the situation remains so
confusing that one might better keep all possibilities open. In
particular, current cosmological data cannot be used as an argument
against the existence of light sterile neutrinos \cite{Hamann2} and
the latter might not necessarily be relevant to current neutrino
oscillation experiments. This mild standpoint motivates us to
examine possible effects of light sterile neutrinos in a neutrino
capture process to directly detect the C$\nu$B.

We find that such light sterile neutrinos can leave a distinct
imprint on the electron energy spectrum in the capture of relic
electron neutrinos by means of radioactive beta-decaying nuclei.
Considering both the (3 + 1) and (3 + 2) schemes of neutrino mixing,
we calculate possible signals of sterile neutrinos relative to
active neutrinos, characterized by their masses and sensitive to
their number densities, in the reaction $\nu^{}_e + ~^3{\rm H} \to
~^3{\rm He} + e^-$ against the corresponding tritium beta decay.
Although our numerical results are just for the purpose of
illustration, we stress that this kind of direct laboratory
detection of the C$\nu$B and its sterile component might not be
hopeless in the long term.

\vspace{0.3cm}

\framebox{\large\bf 2} ~ In the presence of $N^{}_s$ species of
light sterile neutrinos, the flavor eigenstates of three active
neutrinos can be written as
\begin{equation}
\left|{\nu^{}_\alpha}\right\rangle = \sum_{i} V^*_{\alpha i} \left|
{\nu^{}_i} \right\rangle \; ,
\end{equation}
where $\alpha$ runs over $e$, $\mu$ and $\tau$, $\nu^{}_i$ is a mass
eigenstate of active (for $1 \leq i \leq 3$) or sterile (for $4 \leq
i \leq 3 + N^{}_s$) neutrinos, and $V^{}_{\alpha i}$ stands for an
element of the $3\times (3+ N^{}_s)$ neutrino mixing matrix. For
simplicity, we assume that the light sterile neutrinos under
consideration do not significantly affect the values of two
mass-squared differences and three mixing angles of active neutrinos
extracted from current experimental data on solar, atmospheric,
reactor and accelerator neutrino oscillations \cite{PDG}. In this
assumption we shall use $\Delta m^2_{21} \approx 7.6 \times 10^{-5}
~{\rm eV}^2$ and $|\Delta m^2_{31}| \approx 2.4 \times 10^{-3} ~{\rm
eV}^2$ together with $\theta^{}_{12} \approx 34^\circ$ and
$\theta^{}_{13} \approx 10^\circ$ as typical inputs in our numerical
estimates. Depending on the sign of $\Delta m^2_{31}$, two mass
patterns of three active neutrinos are possible:
\begin{itemize}
\item     the normal hierarchy
$m_1 < m_2 = \sqrt{m^2_1 + \Delta m^2_{21}} < m_3 = \sqrt{m^2_1 +
|\Delta m_{31}^2|} ~$;

\item     the inverted hierarchy
$m^{}_3 < m_1 = \sqrt{m_3^2+|\Delta m_{31}^2|} < m_2 = \sqrt{m_3^2 +
|\Delta m_{31}^2| + \Delta m_{21}^2} ~$.
\end{itemize}
In either case the absolute mass scale ($m^{}_1$ or $m^{}_3$) is
unknown, but its upper bound is expected to be of ${\cal O}(0.1)$ eV
as constrained by current cosmological data \cite{WMAP10}. Following
Ref. \cite{Hamann2}, we assume the masses of sterile neutrinos
($m^{}_4$, $m^{}_5$, $\cdots$) to lie in the sub-eV range. Their
mixing with active neutrinos is constrained by current neutrino
experiments and cosmological data and should be at most of ${\cal
O}(0.1)$ \cite{Karagiorgi,Maltoni}. To illustrate the effect of
sterile neutrinos in a neutrino capture process, we shall simply
take $\theta^{}_{1i} \approx 10^\circ$ (for $i \geq 4$) in our
numerical estimates. So we have $|V^{}_{e1}| \approx 0.804$,
$|V^{}_{e2}| \approx 0.542$, $|V^{}_{e3}| \approx 0.171$ and
$|V^{}_{e4}| \approx 0.174$ in the (3 + 1) scheme; or $|V^{}_{e1}|
\approx 0.792$, $|V^{}_{e2}| \approx 0.534$, $|V^{}_{e3}| \approx
0.168$, $|V^{}_{e4}| \approx 0.171$ and $|V^{}_{e5}| \approx 0.174$
in the (3 + 2) scheme. We reiterate that these numerical inputs are
mainly for the purpose of illustration.

Let us concentrate on the relic neutrino capture reaction $\nu^{}_e
+ ~^3{\rm H} \to ~^3{\rm He} + e^-$, since its corresponding beta
decay $^3{\rm H} \to ~^3{\rm He} + e^- + \overline{\nu}^{}_e$ is
being precisely measured in the KATRIN experiment \cite{Weinheimer}.
The capture rate for each neutrino mass eigenstate $\nu^{}_i$ hidden
in the $\nu^{}_e$ state is given by \cite{Vogel,Blennow}
\begin{equation}
{\cal N}^{(i)}_{\rm C \nu B} \approx 6.5 \,\zeta^{}_i \,
|V^{}_{ei}|^2~{\rm yr}^{-1} \, {\rm MCi}^{-1} \; ,
\end{equation}
in which $\zeta^{}_i \equiv n^{}_{\nu^{}_i}/\langle n^{}_{\nu^{}_i}
\rangle$ denotes the ratio of the number density of relic $\nu^{}_i$
neutrinos around the Earth to its average value in the Universe. The
standard Big Bang model predicts $\langle n^{}_{\nu^{}_i} \rangle
\approx \langle n^{}_{\overline{\nu}^{}_i} \rangle \approx 56 ~{\rm
cm}^{-3}$ today for each species of active neutrinos, and this
prediction is also expected to hold for each species of light
sterile neutrinos if they could be completely thermalized in the
early Universe
\footnote{The sub-eV sterile neutrinos discussed in Ref.
\cite{Hamann2} and here should be most likely to stay in full
thermal equilibrium in the early Universe, provided their mixing
with active neutrinos is not strongly suppressed \cite{Hannestad}.
We are indebted to S. Hannestad for clarifying this point to us.}.
Although possible interactions and oscillations in the early
Universe could slightly modify the values of $\langle
n^{}_{\nu^{}_i} \rangle$ and $\langle n^{}_{\overline{\nu}^{}_i}
\rangle$, such corrections are actually unimportant for our
numerical estimates \cite{Blennow,Mangano}. In Eq. (2) the unit MCi
measures one megacurie source of tritium (about 100 g or
$2.1\times10^{25}$ atoms). Note that each $\nu^{}_i$ state yields a
monoenergetic electron, whose kinetic energy is given by $T^{i}_{e}
= Q^{}_{\beta} + E^{}_{\nu^{}_i} $ with $Q^{}_\beta = M^{}_{^3{\rm
H}} - M^{}_{^3{\rm He}} - m^{}_e \approx 18.6$ keV for tritium.
Because at least two active neutrinos are already non-relativistic
today in the C$\nu$B (i.e., their masses are much larger than
$T^{}_\nu \approx 1.945$ K or $\langle p^{}_\nu \rangle = 3 T^{}_\nu
\approx 5 \times 10^{-4}$ eV) and the involved sterile neutrinos are
assumed to have masses of ${\cal O}(0.1)$ eV, we arrive at
$T^{i}_{e} \approx Q^{}_\beta + m^{}_{i}$ as a good approximation.
If the mass of the lightest active neutrino is below $\langle
p^{}_\nu \rangle$ (i.e., it remains hot or relativistic today), then
its energy can be expressed as $E^{}_{\nu^{}_i} \approx
\sqrt{\langle p^{}_\nu \rangle^2 + m^2_i}$, which is of ${\cal
O}(\langle p^{}_\nu \rangle)$ and hence has little effect on the
overall electron energy spectrum. Given a finite energy resolution
in practice, the ideally discrete energy lines of the electrons
emitted from the reaction $\nu^{}_e + ~^3{\rm H} \to ~^3{\rm He} +
e^-$ must spread and form a continuous spectrum. As usual, we
consider a Gaussian energy resolution function defined by
\begin{equation}
R(T^{}_{e}, \, T^{i}_{e}) = \frac{1}{\sqrt{2\pi} \,\sigma}
\exp\left[-\frac{(T^{}_{e} - T^{i}_{e})^2}{2\sigma^2} \right] \; ,
\end{equation}
where $T^{}_{e}$ is the overall kinetic energy of the electrons
detected in the experiment. Using $\Delta$ to denote the
experimental energy resolution (i.e., the full width at half maximum
of a Gaussian energy resolution for the outgoing electrons
\cite{Vogel,Blennow}), we have $\Delta = 2\sqrt{2\ln 2} \,\sigma
\approx 2.35482 \,\sigma$. Then the overall neutrino capture rate
(i.e., the energy spectrum of the detected electrons for the
reaction $\nu^{}_e + ~^3{\rm H} \to ~^3{\rm He} + e^-$) is given as
\begin{equation}
{\cal N}^{}_{\rm C \nu B} = \sum_i {\cal N}^{(i)}_{\rm C \nu B} \,
R(T^{}_{e}, \, T^{i}_{e}) \approx 6.5 \,\sum_i \zeta_i \,|V_{ei}|^2
\, R(T^{}_{e}, \, T^{i}_{e}) ~{\rm yr}^{-1} \, {\rm MCi}^{-1} \; .
\end{equation}
Taking account of the gravitational clustering of relic
non-relativistic neutrinos around the Earth \cite{Wong}, we expect
$\zeta^{}_i \geq 1$ in general. For simplicity, we shall first
assume $\zeta^{}_i =1$ in our numerical estimates of the capture
rate and then illustrate a possible enhancement of the signal due to
much larger values of $\zeta^{}_i$.

The main background of the neutrino capture process under
consideration is the standard tritium beta decay $^3{\rm H} \to
~^3{\rm He} + e^- + \overline{\nu}^{}_e$. In this process the effect
induced by nonzero neutrino masses can show up near the electron's
endpoint energy $Q^{}_\beta - {\rm min}(m^{}_i)$, where ${\rm
min}(m^{}_i)$ means the lightest neutrino mass among $m^{}_i$ (for
$i =$1, 2, $\cdots$, $3 + N^{}_s$). The finite energy resolution may
push the above endpoint towards a higher energy region, and hence it
is likely to mimic the desired signal of the neutrino capture
reaction. Given a finite energy resolution described by the Gaussian
function in Eq. (3), the energy spectrum of the tritium beta decay
can be expressed as
\begin{eqnarray}
~~~~~\frac{{\rm d} {\cal N}^{}_\beta}{{\rm d}T^{}_e} =
\int_0^{Q^{}_{\beta}- {\rm min}(m^{}_i)} {\rm d} T^\prime_e \, \left\{
N^{}_{\rm T} \, \frac{G^2_{\rm F} \, \cos^2\theta^{}_{\rm C}}{2\pi^3}
\, F(Z,\, E^{}_{e}) \, |{\cal M}|^2 \,
\sqrt{E^2_e - m^2_e} \, E^{}_{e}(Q^{}_{\beta}
- T^\prime_e) \right .
\nonumber \\
\left . \times \sum_i\left[ |V^{}_{ei}|^2\sqrt{\left(Q^{}_{\beta}-
T^\prime_e \right)^2 - m_i^2} ~ \Theta(Q^{}_{\beta} - T^\prime_e -
m^{}_i) \right] R(T^{}_e, \, T^\prime_e) \right\} \; ,
~~~~~~~~~~\,\,
\end{eqnarray}
where $E^{}_{e}=T^\prime_e + m^{}_e$ is the electron energy with
$T^\prime_e$ being its kinetic component, $N^{}_{\rm T} \approx 2.1
\times 10^{25}$ denotes the target factor whose value is equal to
the number of tritium atoms of the target, $F(Z, E^{}_{e})$
represents the Fermi function, and $|{\cal M}|^2 \approx 5.55$
stands for the dimensionless contribution of relevant nuclear matrix
elements \cite{Weinheimer}. In Eq. (5) the theta function
$\Theta(Q^{}_{\beta} - T^\prime_e - m^{}_i)$ is adopted to ensure
the kinematic requirement.

\vspace{0.3cm}

\framebox{\large\bf 3} ~ With the help of Eqs. (4) and (5), one may
numerically calculate the relic neutrino capture rate against the
corresponding tritium beta decay. Blennow has done such an analysis
in the scheme of three active neutrinos by examining the dependence
of both the C$\nu$B signal and its background on the neutrino mass
hierarchy and neutrino mixing angles \cite{Blennow}. Here we focus
on the sterile component of the C$\nu$B and its possible signals in
the neutrino capture reaction $\nu^{}_e + ~^3{\rm H} \to ~^3{\rm He}
+ e^-$. As mentioned in section 2, we are mainly interested the
sub-eV sterile neutrinos whose mixing with active neutrinos is of
${\cal O}(0.1)$ or somewhat smaller. To be more explicit, we assume
that the masses of sterile neutrinos are larger than the absolute
mass scale of three active neutrinos. In this case we have the
following naive expectations:
\begin{itemize}
\item     The signal of the sterile component of the C$\nu$B is on
the right-hand side of the electron energy spectrum as compared with
the signal of the active component of the C$\nu$B. Their separation
is measured by their mass differences.

\item     The rate of events for the signal of relic
sterile neutrinos is crucially dependent on the magnitude of
their mixing with active neutrinos. Hence a larger value of
$|V^{}_{ei}|$ (for $i\geq 4$) leads to a higher rate of signal
events.

\item     Whether a signal can be separated from its background
depends on the finite energy resolution $\Delta$ in a realistic
experiment. In general, $\Delta \leq m^{}_i/2$ (for $i=1, 2, \cdots,
3 + N^{}_s$) is required to detect the C$\nu$B via a neutrino
capture reaction \cite{Vogel,Blennow}.
\end{itemize}
Let us make some quantitative estimates of the C$\nu$B signals in
two schemes of neutrino mixing: (a) the (3 + 1) scheme with one
sterile neutrino; and (b) the (3 + 2) scheme with two sterile
neutrinos. In either case the lightest active neutrino mass is
typically taken to be 0.0 eV, 0.05 eV or 0.1 eV. We fix $m^{}_4
=0.3$ eV together with $|V^{}_{e1}| \approx 0.804$, $|V^{}_{e2}|
\approx 0.542$, $|V^{}_{e3}| \approx 0.171$ and $|V^{}_{e4}| \approx
0.174$ in the (3 + 1) scheme; or $m^{}_4 = 0.2$ eV and $m^{}_5 =
0.4$ eV together with $|V^{}_{e1}| \approx 0.792$, $|V^{}_{e2}|
\approx 0.534$, $|V^{}_{e3}| \approx 0.168$, $|V^{}_{e4}| \approx
0.171$ and $|V^{}_{e5}| \approx 0.174$ in the (3 + 2) scheme. The
gravitational clustering of relic neutrinos around the Earth is
tentatively omitted and will be illustrated later. Our numerical
results are presented in Figs. 1---3. Some discussions are in order.

Fig. 1 shows the relic neutrino capture rate as a function of the
kinetic energy $T^{}_e$ of electrons in the (3 + 1) scheme with
$\Delta m^2_{31} >0$ and $m^{}_4 = 0.3$ eV. The value of the finite
energy resolution $\Delta$ is taken in such a way that only the
signal of the sterile neutrino can be seen (left panel) or both the
signals of active and sterile neutrinos can be seen (right panel).
These simple results confirm the naive expectations given above,
especially for the signal of the sterile component of the C$\nu$B.
We find that it is in principle possible to distinguish the sterile
neutrino from the background when $\Delta$ is of ${\cal O}(0.1)$ eV
or much smaller. As the lightest neutrino mass $m^{}_1$ increases
from 0 to 0.1 eV, the signal curve moves towards the higher $T^{}_e
- Q^{}_\beta$ region while the background curve moves towards the
lower $T^{}_e - Q^{}_\beta$ region. To make a realistic measurement
sensitive to the active component of the C$\nu$B, one needs a
sufficiently good energy resolution. The required energy resolution
depends on the mass hierarchy of three active neutrinos, as one can
see from Fig. 1 (right panel). Note that the small peak sitting at
$T^{}_e - Q^{}_\beta \approx 0.05$ eV in the top right corner of
Fig. 1 arises from the contribution of $\nu^{}_3$ with $m^{}_3
\approx 0.05$ eV. It is not seeable in the top left corner of Fig. 1
simply because the resolution is not good enough.

An analogous analysis of the relic neutrino capture rate is carried
out in the (3 + 1) scheme with $\Delta m^2_{31} < 0$ and $m^{}_4 =
0.3$ eV, and the numerical results are given in Fig. 2. We see that
this figure is essentially similar to Fig. 1, and the primary
difference appears in the signal peaks of active neutrinos in their
top right corners. Such a difference can be understood with the help
of Eqs. (3) and (4): the central position of a signal peak is
approximately determined by $T^{}_e - Q^{}_\beta \approx
E^{}_{\nu^{}_i}$ and its height is mainly measured by
$|V^{}_{ei}|^2$ for given values of $\Delta$ and $\zeta^{}_i$. When
the lightest neutrino mass $m^{}_3$ is extremely small, the main
contribution of active neutrinos to the capture rate sits at $T^{}_e
- Q^{}_\beta \approx m^{}_1 \approx m^{}_2 \approx 0.05$ eV as shown
in the top right corner of Fig. 2. In contrast, the main signal
peaks shows up at $T^{}_e - Q^{}_\beta \in \{m^{}_1, \, m^{}_2\}$
when the lightest neutrino mass $m^{}_1$ is extremely small (as
illustrated in the top right corner of Fig. 1). So it is in practice
much easier to detect the active component of the C$\nu$B when three
active neutrinos have an inverted mass hierarchy or a nearly
degenerate mass spectrum.

We illustrate the relic neutrino capture rate in the (3 + 2) scheme
with $\Delta m^2_{31} > 0$ (left panel) or $\Delta m^2_{31} < 0$
(right panel) in Fig. 3, where $m^{}_4 = 0.2$ eV and $m^{}_5 = 0.4$
eV are typically taken. The value of the finite energy resolution
$\Delta$ is chosen in such a way that only the signals of sterile
neutrinos can be seen. If the values of $m^{}_4$ and $m^{}_5$ are
very close to each other or their difference is much smaller than
$\Delta$, it will be very difficult to distinguish between the two
signal peaks of sterile neutrinos. When the absolute mass scale of
three active neutrinos is very close to the smaller mass of two
sterile neutrinos, there may be a mixture between their signals as
shown in the bottom left and bottom right corners of Fig. 3. In this
case only the heavier sterile neutrino is distinguishable from the
signal of active neutrinos.

So far we have neglected possible gravitational clustering of relic
neutrinos in our numerical estimates. To illustrate this effect, we
calculate the relic neutrino capture rate in the (3 + 2) scheme with
$m^{}_1 =0$ or $m^{}_3 =0$, $m^{}_4 =0.2$ eV and $m^{}_5 =0.4$ eV.
We assume a larger gravitational clustering effect for a heavier
neutrino around the Earth \cite{Wong}. In this reasonable assumption
we typically take $\zeta^{}_1 = \zeta^{}_2 = \zeta^{}_3 =1$ (without
clustering effects for three active neutrinos because their maximal
mass is about 0.05 eV in the scenario under discussion) and
$\zeta^{}_5 = 2 \zeta^{}_4 = 10$ (with mild clustering effects for
two sterile neutrinos because their masses are 0.2 eV and 0.4 eV,
respectively). The value of the finite energy resolution $\Delta$ is
chosen in such a way that only the signals of sterile neutrinos can
be observed. As shown in Fig. 4, the signals of two sterile
neutrinos are obviously enhanced due to $\zeta^{}_4 >1$ and
$\zeta^{}_5 >1$. If the gravitational clustering of non-relativistic
neutrinos is very significant around the Earth, it will be very
helpful for us to detect the C$\nu$B by means of the neutrino
capture processes.

\vspace{0.3cm}

\framebox{\large\bf 4} ~ Motivated by a mild standpoint that current
cosmological data cannot be used as an argument against the
existence of light sterile neutrinos and the latter might not
necessarily be very relevant to current neutrino oscillation
experiments, we have examined possible effects of sub-eV sterile
neutrinos in a neutrino capture reaction to directly detect the
C$\nu$B. We find that such light sterile neutrinos can in principle
leave a distinct imprint on the electron energy spectrum in the
capture of relic electron neutrinos by means of radioactive
beta-decaying nuclei. Considering both the (3 + 1) and (3 + 2)
schemes of neutrino mixing, we have calculated possible signals of
sterile neutrinos relative to active neutrinos in the neutrino
capture process $\nu^{}_e + ~^3{\rm H} \to ~^3{\rm He} + e^-$
against the corresponding tritium beta decay. We reiterate that our
numerical results are just for the purpose of illustration, but they
are useful to show a few salient features of the active and sterile
components of the C$\nu$B in such neutrino capture processes.

The major limiting factor of a neutrino capture experiment is its
energy resolution. If the energy resolution is gradually improved,
it should be possible to establish a signal of the C$\nu$B
beyond the endpoint of the electron energy spectrum of the beta decay
in the future. One might only be able to observe one peak for a
given energy resolution, either because the peaks of different
neutrino mass eigenstates merge into a wide one or because the peaks
of lighter neutrinos are overwhelmed by the background (or for both reasons). In this case some information on the sterile component
of the C$\nu$B could be obtained from a combined analysis of the
available data from the neutrino capture experiment, neutrino
oscillations and cosmological observations. If multiple peaks
could finally be seen in the electron energy spectrum with a
sufficiently good energy resolution, then they would convincingly 
indicate the presence of light sterile neutrinos in the C$\nu$B.

Needless to say, a direct detection of the C$\nu$B in a realistic
experiment is very complicated and sophisticated. We admit that
current experimental techniques are unable to lead us to a
guaranteed detection of the C$\nu$B, but more and more efforts have
been being made towards this ultimate goal. Whether the neutrino
capture experiment (e.g., $\nu^{}_e + ~^3{\rm H} \to ~^3{\rm He} + e^-$
under discussion) can win the race or not remains an open question,
but it is certainly the most promising horse at present.

\vspace{0.6cm}

One of us (Z.Z.X.) would like to thank S. Hannestad for a very 
helpful discussion. This work is supported in part by the National Natural
Science Foundation of China under grant No. 10425522 and No. 10875131.

\newpage

\begin{figure}[p!]
\begin{center}
\begin{tabular}{cc}
\includegraphics*[bb=17 15 268 210, width=0.46\textwidth]{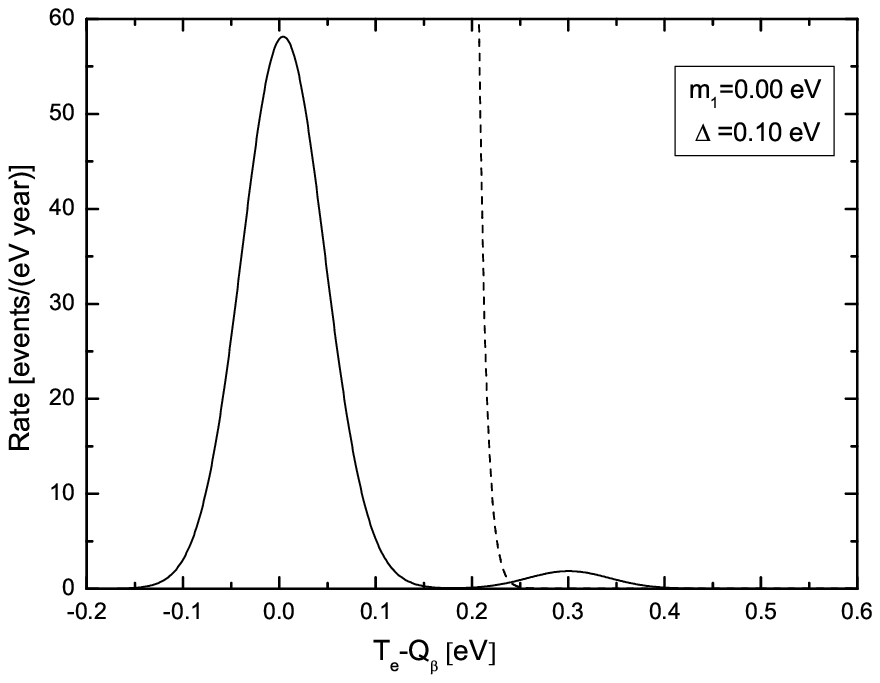}
&
\includegraphics*[bb=18 16 268 208, width=0.46\textwidth]{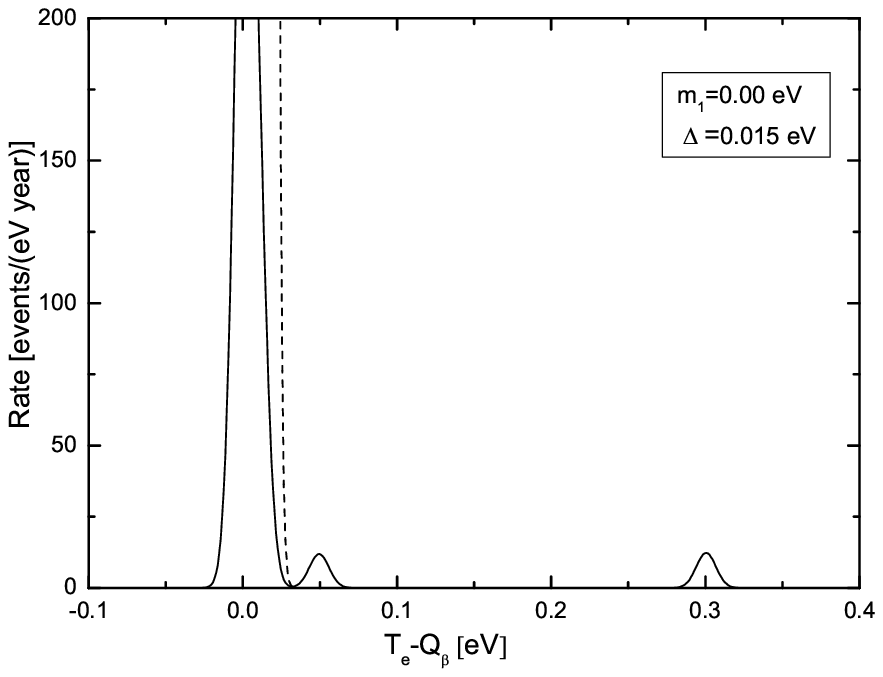}
\\
\includegraphics*[bb=17 15 268 210, width=0.46\textwidth]{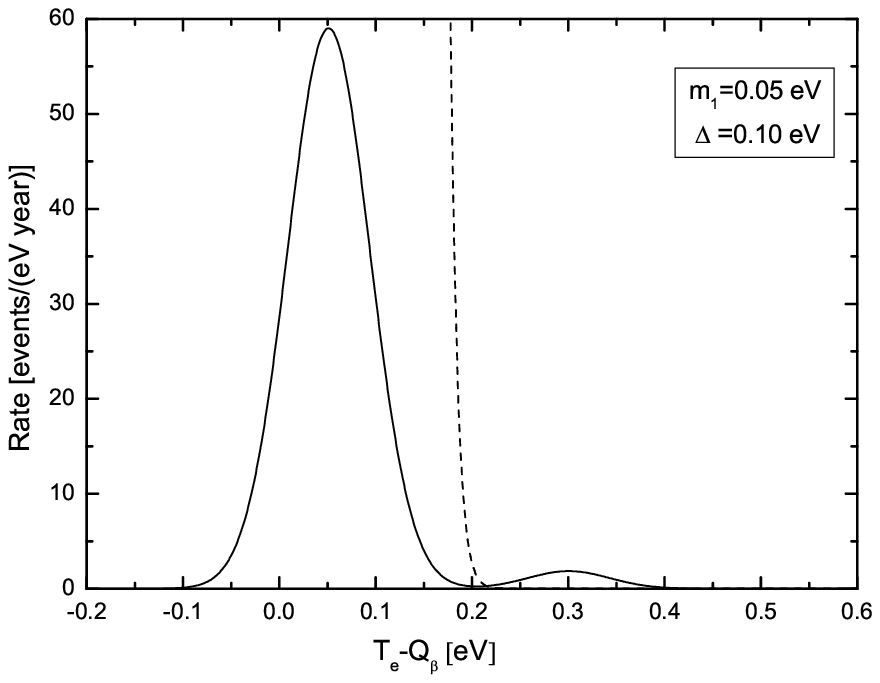}
&
\includegraphics*[bb=18 16 268 208, width=0.46\textwidth]{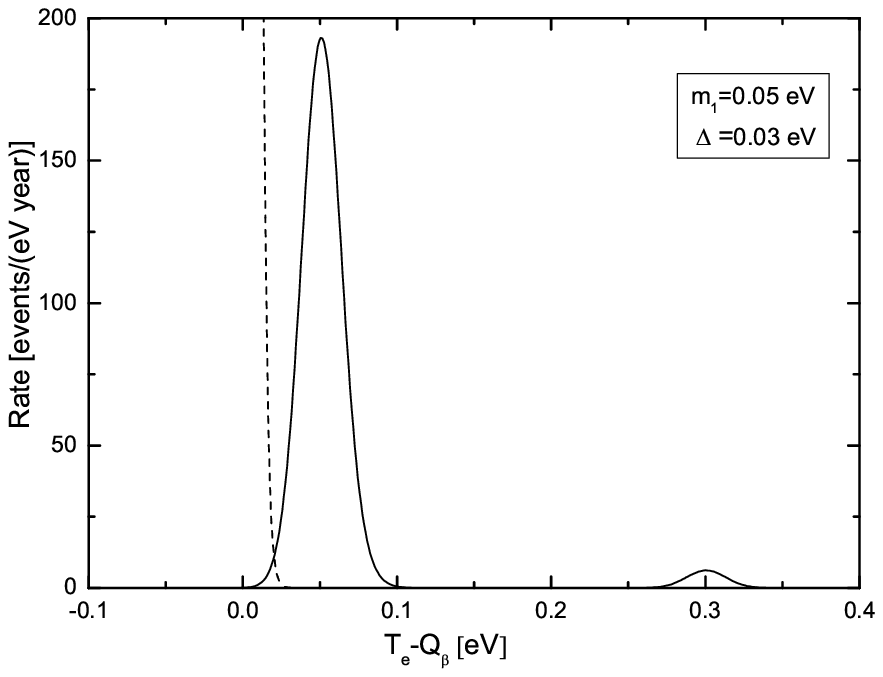}
\\
\includegraphics*[bb=17 15 268 210, width=0.46\textwidth]{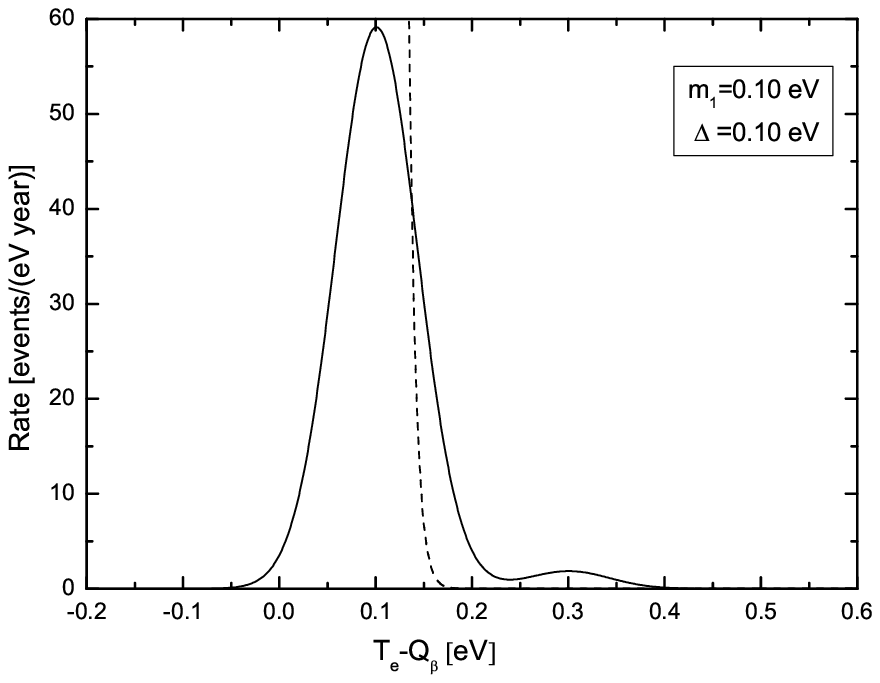}
&
\includegraphics*[bb=18 16 268 208, width=0.46\textwidth]{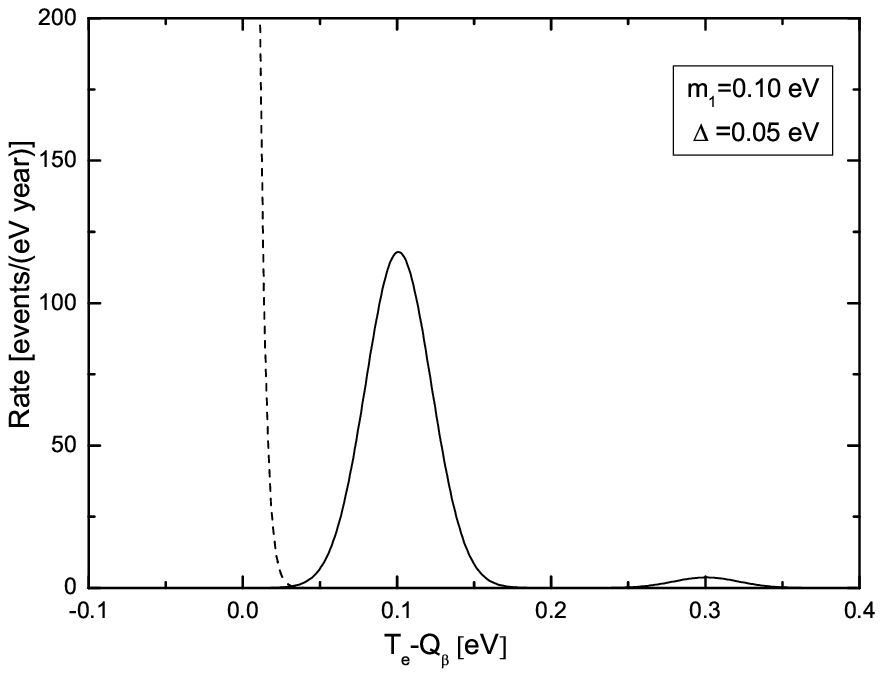}
\end{tabular}
\end{center}
\caption{The relic neutrino capture rate as a function of the
kinetic energy $T^{}_e$ of electrons in the (3 + 1) scheme with
$\Delta m^2_{31} >0$ and $m^{}_4 = 0.3$ eV. The solid and dashed
curves represent the C$\nu$B signal and its background,
respectively. The value of the finite energy resolution $\Delta$ is
taken in such a way that only the signal of the sterile neutrino can
be seen (left panel) or both the signals of active and sterile
neutrinos can be seen (right panel). The gravitational clustering of
relic neutrinos around the Earth has been omitted.}
\end{figure}

\begin{figure}[p!]
\begin{center}
\begin{tabular}{cc}
\includegraphics*[bb=17 15 268 210, width=0.46\textwidth]{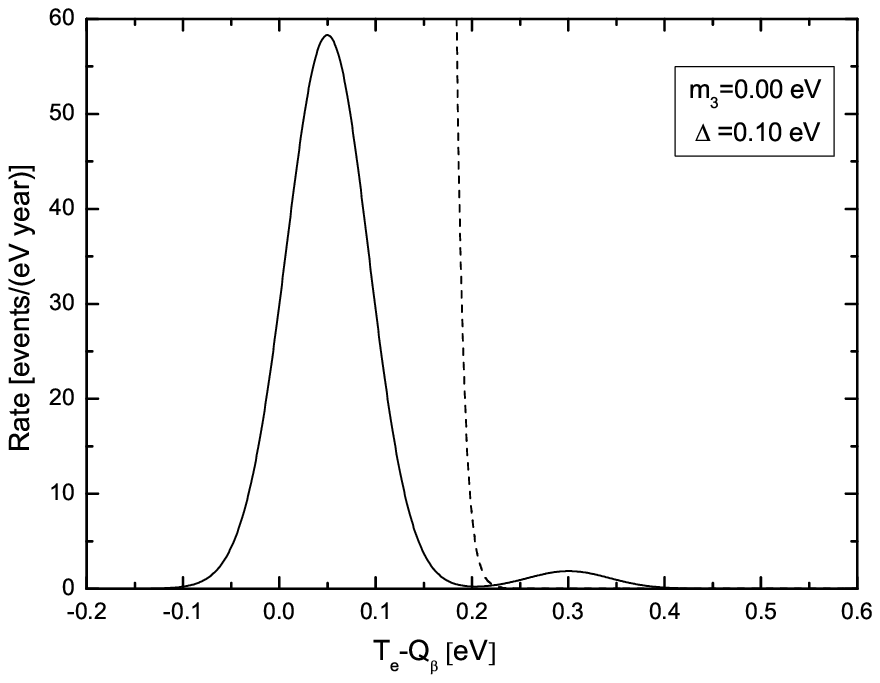}
&
\includegraphics*[bb=18 16 268 208, width=0.46\textwidth]{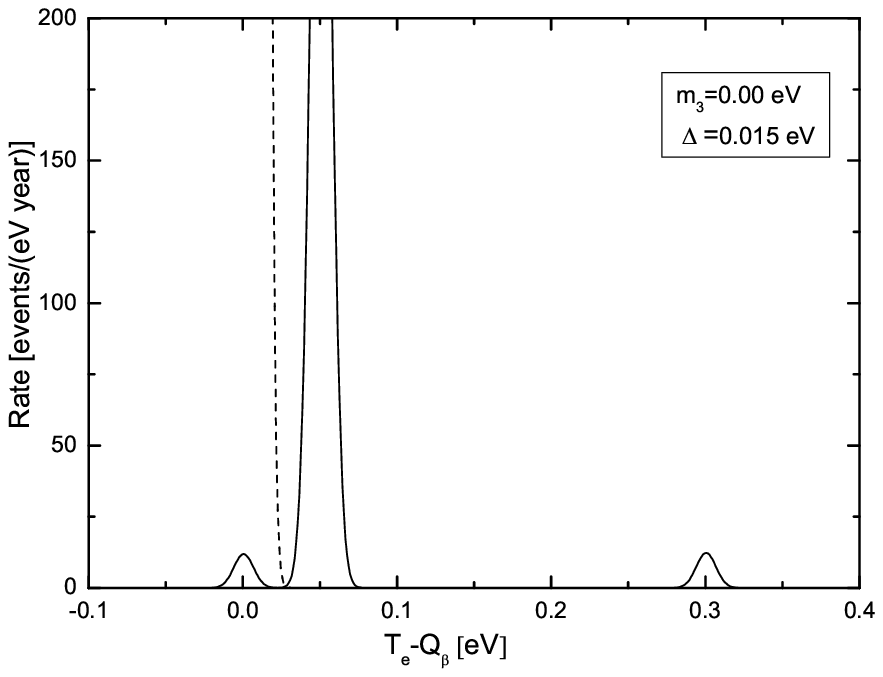}
\\
\includegraphics*[bb=17 15 268 210, width=0.46\textwidth]{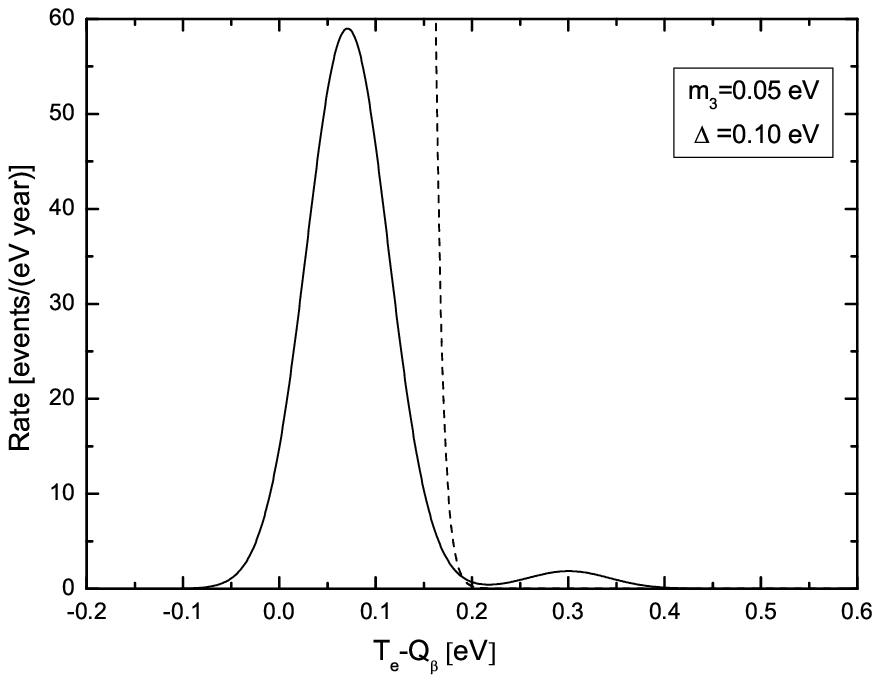}
&
\includegraphics*[bb=18 16 268 208, width=0.46\textwidth]{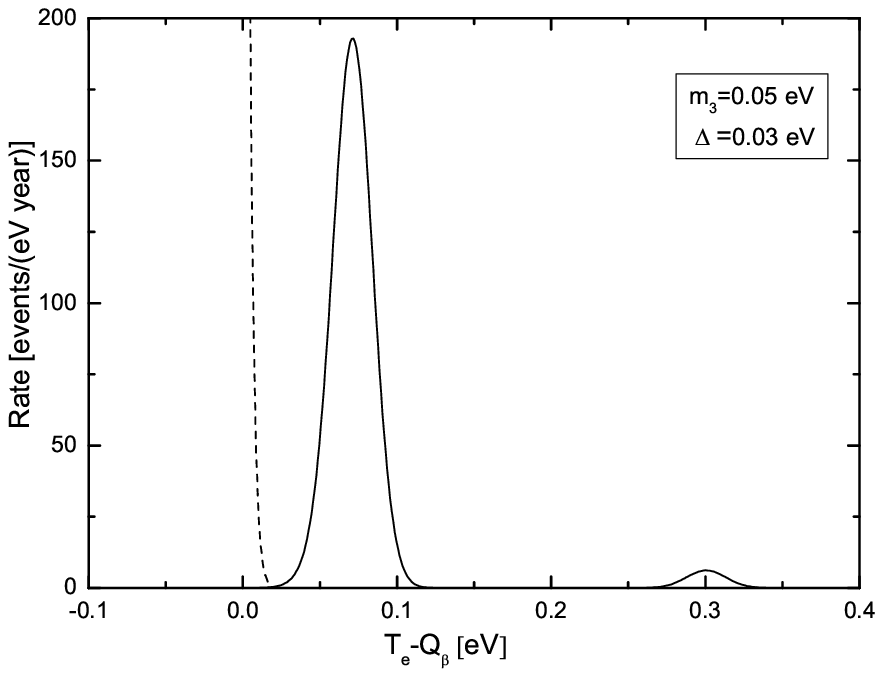}
\\
\includegraphics*[bb=17 15 268 210, width=0.46\textwidth]{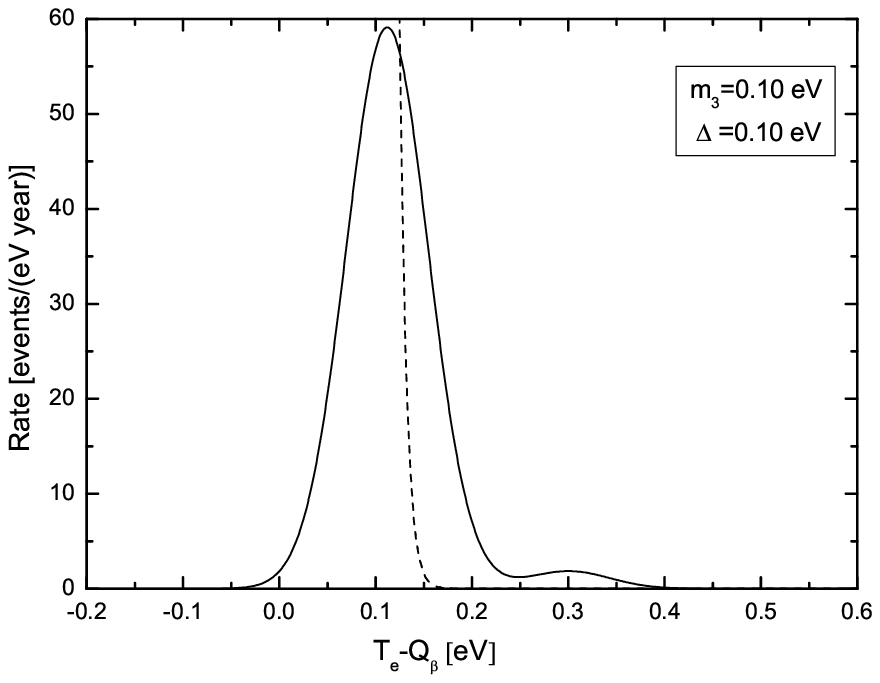}
&
\includegraphics*[bb=18 16 268 208, width=0.46\textwidth]{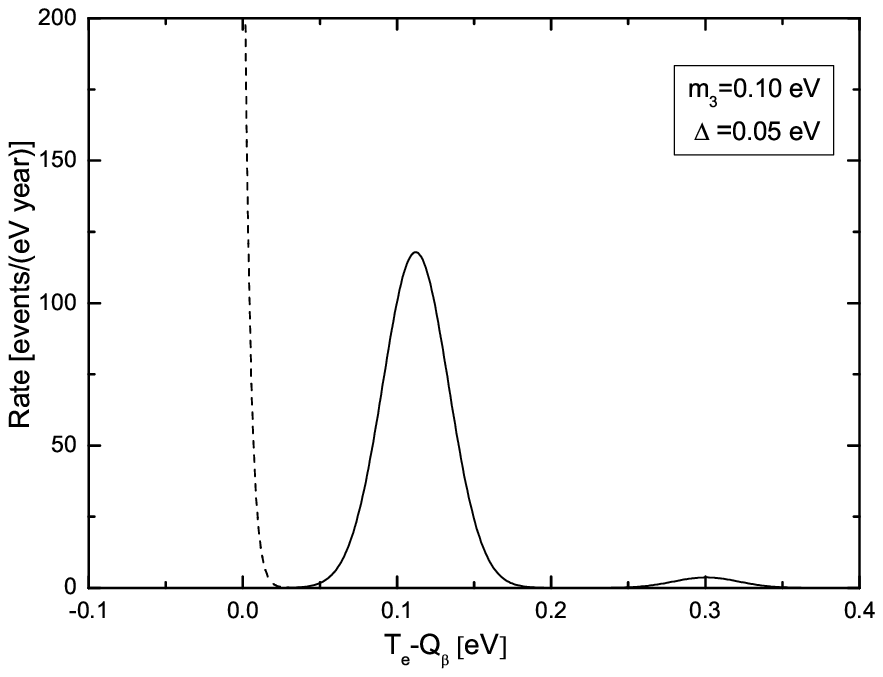}
\end{tabular}
\end{center}
\caption{The relic neutrino capture rate as a function of the
kinetic energy $T^{}_e$ of electrons in the (3 + 1) scheme with
$\Delta m^2_{31} <0$ and $m^{}_4 = 0.3$ eV. The solid and dashed
curves represent the C$\nu$B signal and its background,
respectively. The value of the finite energy resolution $\Delta$ is
taken in such a way that only the signal of the sterile neutrino can
be seen (left panel) or both the signals of active and sterile
neutrinos can be seen (right panel). The gravitational clustering of
relic neutrinos around the Earth has been omitted.}
\end{figure}

\begin{figure}[p!]
\begin{center}
\begin{tabular}{cc}
\includegraphics*[bb=17 15 268 210, width=0.46\textwidth]{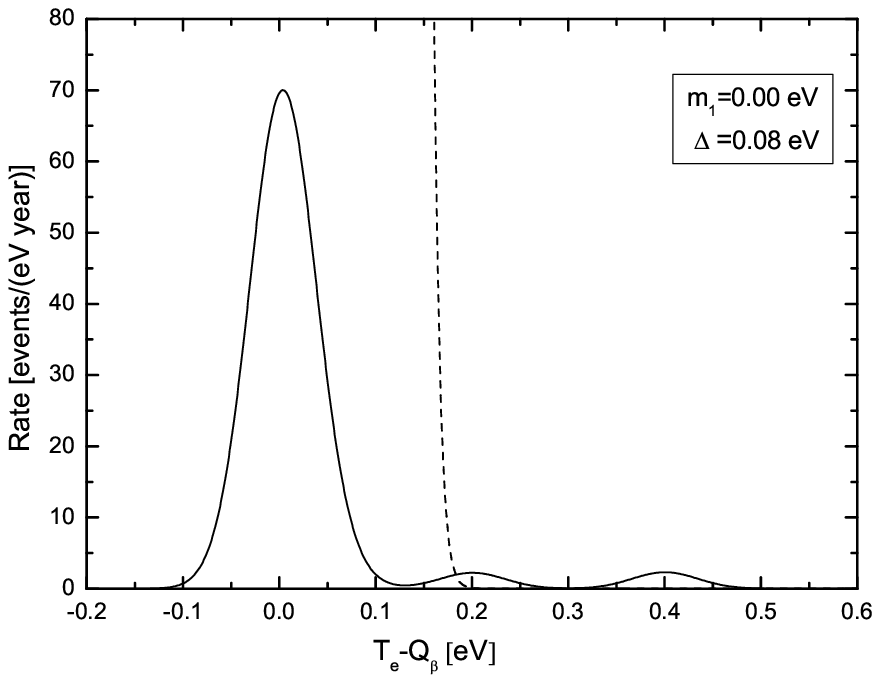}
&
\includegraphics*[bb=17 15 268 210, width=0.46\textwidth]{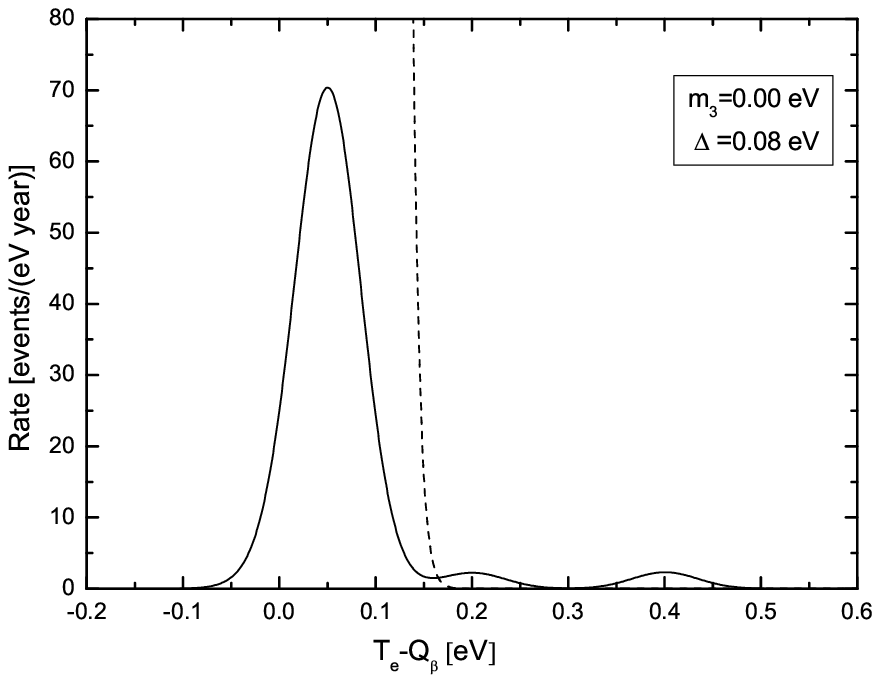}
\\
\includegraphics*[bb=17 15 268 210, width=0.46\textwidth]{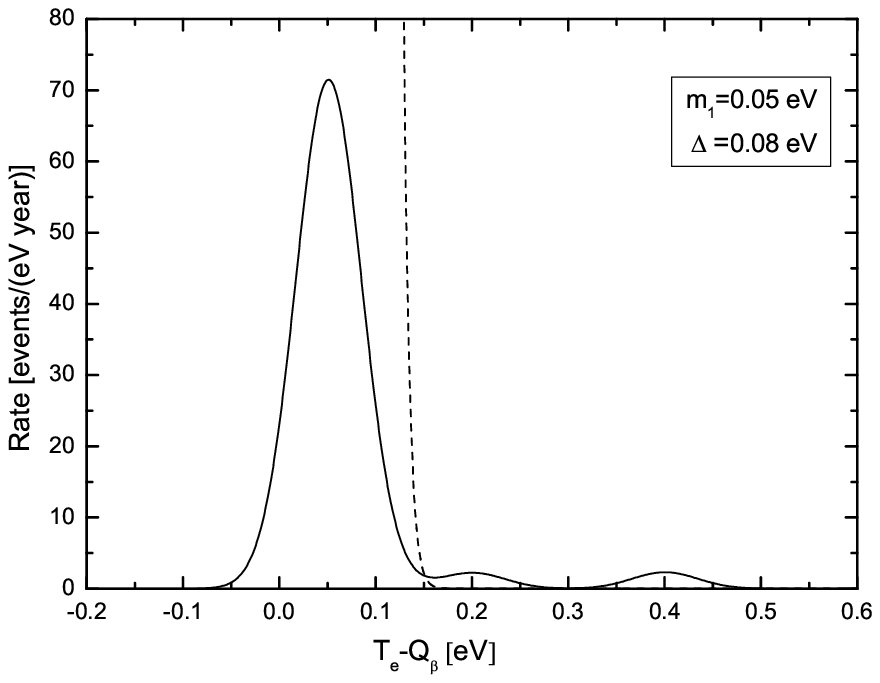}
&
\includegraphics*[bb=17 15 268 210, width=0.46\textwidth]{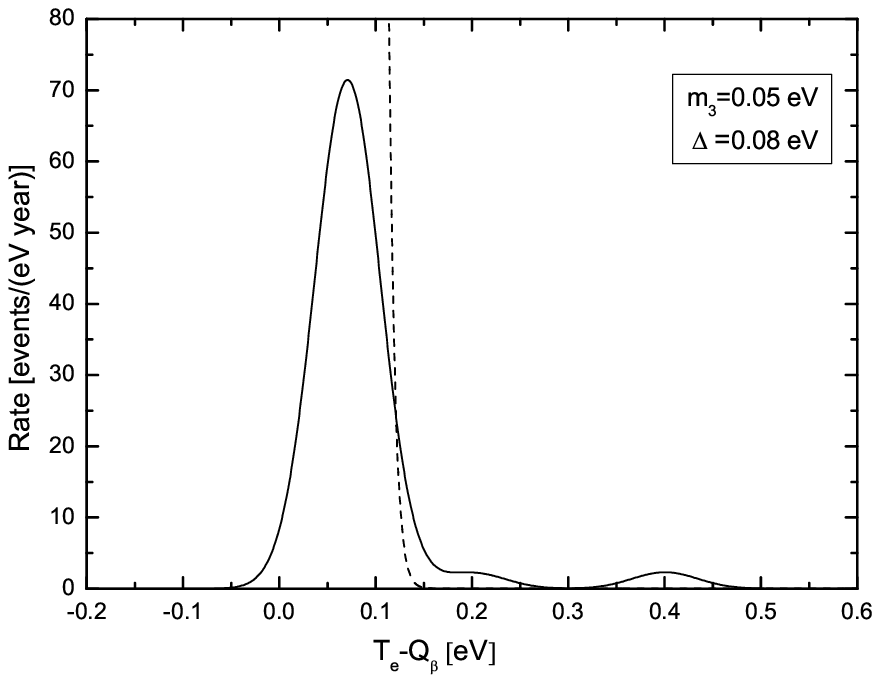}
\\
\includegraphics*[bb=17 15 268 210, width=0.46\textwidth]{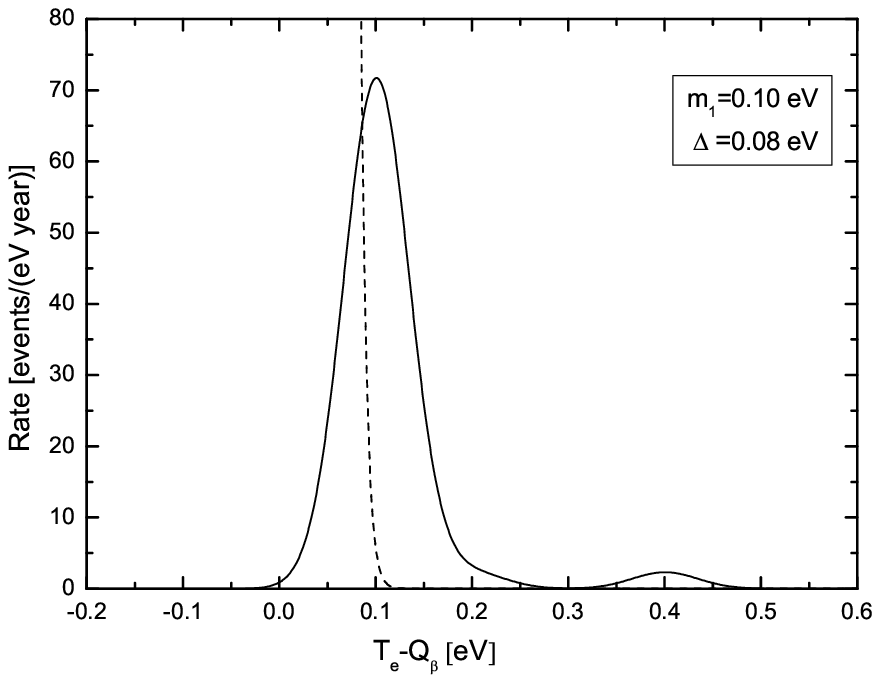}
&
\includegraphics*[bb=17 15 268 210, width=0.46\textwidth]{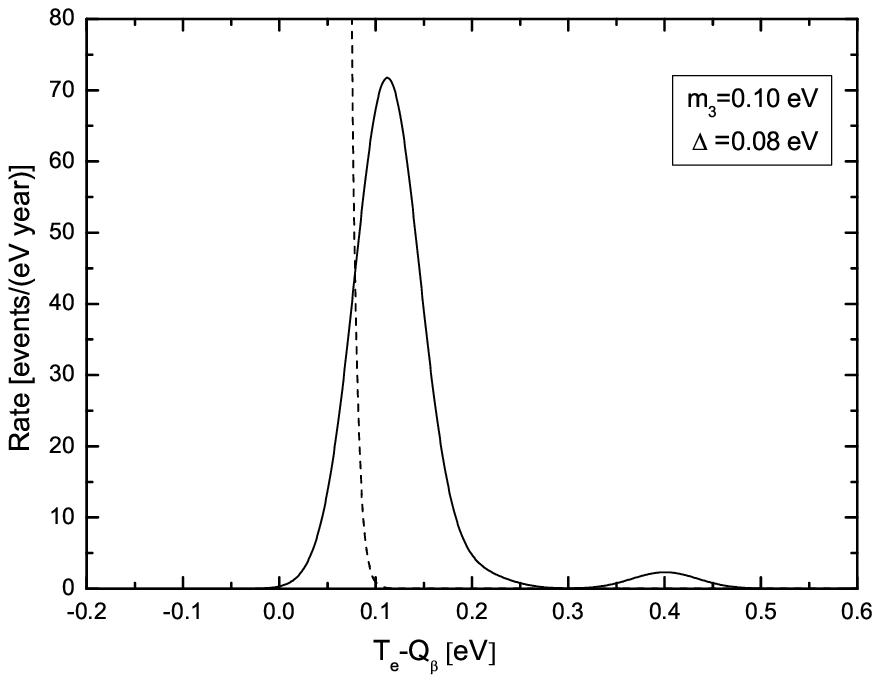}
\end{tabular}
\end{center}
\caption{The relic neutrino capture rate as a function of the
kinetic energy $T^{}_e$ of electrons in the (3 + 2) scheme with
$\Delta m^2_{31} >0$ (left panel) or $\Delta m^2_{31} <0$ (right
panel). In either case $m^{}_4 = 0.2$ eV and $m^{}_5 = 0.4$ eV are
typically taken. The solid and dashed curves represent the C$\nu$B
signal and its background, respectively. The value of the finite
energy resolution $\Delta$ is taken in such a way that only the
signals of sterile neutrinos can be seen. The gravitational
clustering of relic neutrinos around the Earth has been omitted.}
\end{figure}

\begin{figure}[t!]
\begin{center}
\begin{tabular}{cc}
\includegraphics*[bb=17 15 268 210, width=0.46\textwidth]{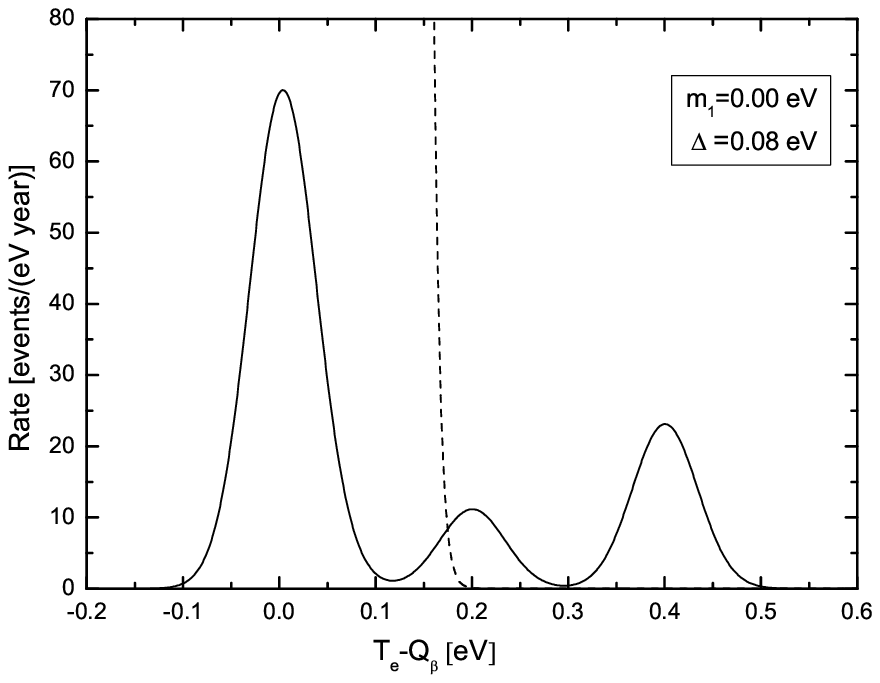}
&
\includegraphics*[bb=17 15 268 210, width=0.46\textwidth]{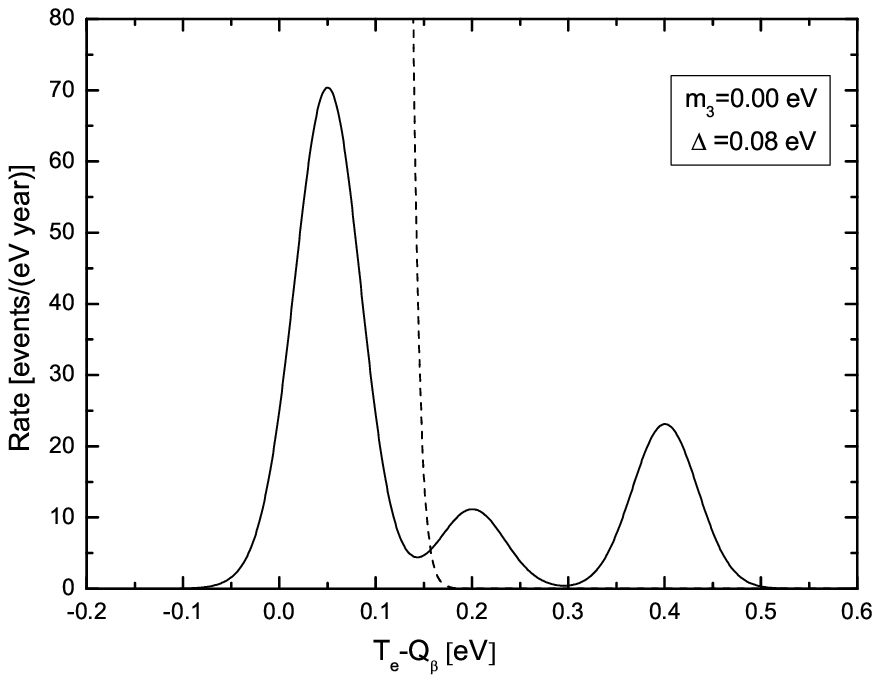}
\end{tabular}
\end{center}
\caption{The relic neutrino capture rate as a function of the
kinetic energy $T^{}_e$ of electrons in the (3 + 2) scheme with
$\Delta m^2_{31} >0$ (left panel) or $\Delta m^2_{31} <0$ (right
panel). In either case $m^{}_4 = 0.2$ eV and $m^{}_5 = 0.4$ eV are
typically taken. The solid and dashed curves represent the C$\nu$B
signal and its background, respectively. The value of the finite
energy resolution $\Delta$ is taken in such a way that only the
signals of sterile neutrinos can be seen. The gravitational
clustering of relic sterile neutrinos around the Earth has been
illustrated by taking $\zeta^{}_1 = \zeta^{}_2 = \zeta^{}_3 =1$ and
$\zeta^{}_5 = 2 \zeta^{}_4 = 10$ for example.}
\end{figure}

\end{document}